\documentclass[twocolumn,showpacs,preprintnumbers,amsmath,amssymb]{revtex4}
\usepackage{graphicx}
\usepackage{dcolumn}
\usepackage{bm}
\begin{document}

\title{Transverse tunneling current through guanine traps in DNA}
\author{Vadim Apalkov$^{\ast\dag}$ and Tapash Chakraborty$^{\ast\ddag}$}
\affiliation{$^\ast$Department of Physics and Astronomy,
University of Manitoba, Winnipeg, Canada R3T 2N2}
\affiliation{$^\dag$Department of Physics and Astronomy, Georgia 
State University, Atlanta, Georgia 30303, USA}
\date{\today}
\begin{abstract}
The current - voltage dependence of the transverse tunneling
current through the electron or hole traps in a DNA is investigated.
The hopping of the charge between the sites of the trap and
the charge-phonon coupling results in a staircase structure of the
I-V curve. For typical parameters of the DNA molecule the energy
characteristics of a DNA trap can be extracted from the I-V dependence,
viz., for a small gate voltage the phonon frequency and for a large 
gate voltage the hopping integral can be found from the positions of 
the steps in the I-V curve. Formation of the polaronic state also 
results in the redistribution of the tunneling current between the 
different sites of the traps.

\end{abstract}
\maketitle
For the past several years intense research on electron (or hole)
migration through DNA has established that these molecules are capable
of transporting charges over a distance of at least a few nanometers
\cite{review,giese,schuster,kelley}. This is primarily achieved by
oxidation of guanines (G) which generates a guanine radical cation. Guanine
has the lowest oxidation potential of the common DNA bases \cite{steenken}
and therefore a guanine radical cation can only oxidize another G. Stacked
Gs such as GG, or a GGG, having much lower ionization potential than that
of an isolated G, known to act as {\it hole traps} \cite{saito}.
Since accumulation of holes at G-sequences usually leads to deleterious
effects, including mutations \cite{boone}, charge migration through DNA
may have important biological consequences. Electronic propreties of both 
electron and hole traps depend crucially on the charge dynamics inside the 
trapping spots. Therefore, it is very important to extract the parameters 
which characterize such dynamics. Here we show how the energy
characteristics of a DNA trap can be extracted from the current (I) -
voltage (V) dependence of the transverse tunneling current through the
electron or hole traps. Hopping of the charge between the sites of the
trap and the charge-phonon coupling results in a staircase structure of
the I-V curve. For a small gate voltage the phonon frequency and for a
large gate voltage the hopping integral can be found from the positions
of the steps in the I-V curve. Formation of the polaronic state also
results in the redistribution of the tunneling current between the
different sites of the traps.

Since our interest is in the local properties of DNA traps, transport in the 
direction perpendicular to the backbone axis (transverse transport \cite{zwolak05}) 
is important. In this case, if the electrodes have a relatively small width 
tunneling occurs through a single DNA base pair. The saturated (unstructured) 
tunneling current then depends on the particular type of base pair \cite{zwolak05}. 
This fact can be used to discover the sequence of DNA by scanning it with 
conducting probes. We demonstrate here that not only the saturated value of the
tunneling current but also the {\it structure} of the I-V curves can provide 
important information on the properties of the DNA, in particular about the 
trapping spots. This is because the tunneling current through the system is 
determined by the density of states (DOS) of the system. For a finite system 
the DOS has peaks corresponding to discrete energy levels. These peaks will 
result in a staircase structure of the tunneling current as a function
of the applied voltage whenever the Fermi levels align with a new state
of the system and thereby open an additional channel for tunneling.
Therefore, from the staircase structure of I-V curve one can learn about
the energy spectra of the system. For DNA the trapping spots consist
of a finite number of base pairs. Hopping between the base pairs within the
traps determine the energy spectra of the spots. In addition to the
energy scale due to hopping there is also an energy scale
due to the electron-phonon (or hole-phonon) ineraction. Finally, for
DNA trapping spots the I-V dependence has two types of staircase
structure, one due to hopping and the other due to the
phonons. We have explored the interplay between these two effects.

The tunneling transport through a single molecule or a
quantum dot with electron-phonon coupling has been extensively
studied in the literature \cite{glazman88,mitra04,galperin04,koch05}.
The main outcome of these  works is the staircase structure of I-V curves
due to phonon sidebands. The heights of the steps in this structure
depend on the strength of the electron-phonon interactions, temperature, and
on the equilibrium condition of the electron-phonon system. These studies
were mainly restricted to a molecule with a single electron
energy level, although a general approach to a many level system is
also formulated \cite{galperin04}.

The DNA trap can be considered as a system of a few molecules (base
pairs) with hopping between them and the electron-phonon coupling.
Then in the I-V curves we should observe the interplay between the
staircase structure due to hopping between the molecules
and due to the phonon sidebands. Since the tunneling occurs only
through a single base pair the I-V structure should also depend on
the position of the base pair through which the tunneling
current is measured. Here we consider only the hole traps and the
tunneling current of holes, but the analysis is also valid for
electron traps and electron transport. Whether it is a hole transport or electron
transport depends on the gate potential, i.e. on the position of the
chemical potential at zero source-drain voltage $V_{sd}$.

We disregard below the effects related to a Coulomb
blockade \cite{coulomb} or to a double occupancy of the DNA traps,
assuming that the repulsion between the holes is strong enough,
although for some range of parameter of the traps the bipolaronic
effect may become strong and a two-hole system can have lower
energy than a single-hole system \cite{traps}. For a single
hole in the trap the Hamiltonian of the trap and the electrodes
consists of three parts: (i) the trap Hamiltonian which include
tight-binding hole part with hopping between the
nearest base pairs, the Holstein's phonon Hamiltonian
with diagonal hole-phonon interaction \cite{holstein}, (ii)
the Hamiltonian of two leads, left (L) and right (R), and
(iii) the Hamiltonian corresponding to tunneling between
the leads and DNA traps
\begin{equation}
{\cal H} = {\cal H}_{trap}+{\cal H}_{leads}+{\cal H}_{t},
\label{H}
\end{equation}
where
\begin{eqnarray}
{\cal H}_{trap}&=&\sum_{i =1}^{N_t}\epsilon\, a^{\dagger}_{i }
a_i-t\sum_i\left[ a^{\dagger}_{i}
a_{i+1}+h.c.\right]+\nonumber \\
&+&\hbar\omega\sum_i b^{\dagger}_i b_i
+\chi\sum_{i} a^{\dagger}_i a_i
\left( b^{\dagger}_i + b_i\right),
\label{Htrap} \\
{\cal H}_{leads}&=&\sum_{k,\alpha=L,R}\epsilon_k\,
d^{\dagger}_{\alpha,k} d_{\alpha,k},
\label{Hi} \\
{\cal H}_t &=&-t_0\sum_{\alpha=L,R,k}
\left[a^{\dagger}_{i_0} d_{\alpha,k}+h.c.\right],
\label{Hph}
\end{eqnarray}
where $a_i$ is the annihilation operator of the hole on site (base
pair) $i$, $\epsilon$ is the on-site energy of the hole in the trap
(same for all base pairs within the trap and is determined by the
gate voltage or doping of DNA), $b_i$ is the annihilation operator of
a phonon on site $i$, $t$ is the hopping integral between the
nearest base pairs, $\omega$ is the phonon frequency,
$\chi$ is hole-phonon coupling constant, and $d_{\alpha,k}$ is the
annihilation operator of a hole in the lead $\alpha=L,R$ with
momentum $k$. The index $i=1,\ldots,N_t$ in Eq.~(\ref{Htrap})
labels the sites (base pairs) in the trap and $N_t$ is their total
number. Tunneling from the leads to the trap occurs only to the
site $i_0$ with the tunneling amplitude $t_0$.
In the hole-phonon part of the DNA Hamiltonian ${\cal H}_{trap}$, we
include only the optical phonons \cite{bishop02} with diagonal
hole-phonon interaction.

\begin{figure}
\begin{center}\includegraphics[width=7.6cm]{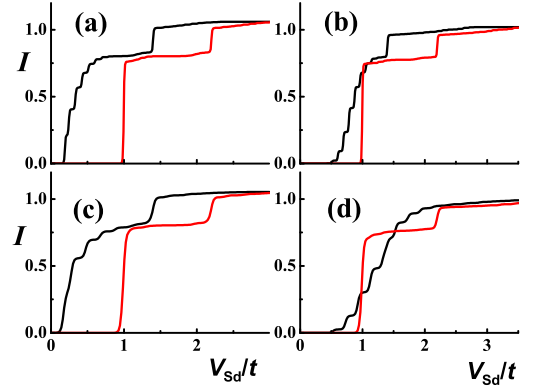}\end{center}
\vspace*{-1cm}
\caption{Current vs the sourse-drain voltage shown for two base pairs
in the trap ($N_t=2$) and different values of phonon frequency and
hole-phonon interaction strength: (a) $\gamma=0.1$, $\lambda
=0.5$; (b)  $\gamma=0.1$, $\lambda=1.0$; (c)  $\gamma=0.2$, $\lambda
=0.5$;  (d)  $\gamma=0.2$, $\lambda=1.0$. Black line corresponds to
$\epsilon= 1.3 t$ while the red line is for $\epsilon=2.7$.
}
\label{fig:Fig1}
\end{figure}

We describe the process of tunneling through the trap as a sequential
tunneling \cite{luryi}.  In the weak lead-trap coupling regime the tunneling
Hamiltonian ${\cal H}_t$ can be considered as a perturbation which
introduce the transitions between the states of the trap Hamiltonian,
${\cal H}_{trap}$. We denote the eigenstates of the
trap Hamiltonian without coupling to the leads as
$\left| 0,m\right\rangle$ with energy $E_{0,m}$ for the trap
without any holes, and $\left| 1,n\right\rangle$ with
the energy $E_{1,n}$ for trap with a single hole.
In the weak lead-trap coupling limit the  master equation for
the density matrix of the trap reduces to the
rate equation \cite{mitra04} for probability $P_{0,m}$ to occupy
the state $\left| 0,m\right\rangle $ and probability
$P_{1,n}$ to occupy the state $\left| 1,n\right\rangle$,
\begin{eqnarray}
\frac{d P_{1,n}}{dt}  &=&
\sum_{m,\alpha=L,R} W_{\alpha,mn}^{0\rightarrow 1} P_{0,m} -
 \sum_{m,\alpha=L,R} W_{\alpha,nm}^{1\rightarrow 0} P_{1,n} -
\nonumber \\
& &-\frac1{\tau}\left[P_{1,n}-P_{1,n}^{eq}
\sum_{n^{\prime}} P_{1,n^{\prime}}
\right],
\label{P1}
\end{eqnarray}
\begin{eqnarray}
\frac{d P_{0,m}}{dt} &=&
\sum_{n,\alpha=L,R} W_{\alpha,nm}^{1\rightarrow 0} P_{1,n} -
 \sum_{n,\alpha=L,R} W_{\alpha,mn}^{0\rightarrow 1} P_{0,m} -
\nonumber \\
 & &-\frac1{\tau} \left[P_{0,m}-P_{0,m}^{eq}
\sum_{m^{\prime}} P_{0,m^{\prime}}\right].
\label{P0}
\end{eqnarray}
The distributions
$P_{1,n}^{eq}$ and $P_{0,m}^{eq}$ are the corresponding equilibrium
distributions with temperature $T$,
$P_{1,n}^{eq} =  \exp\left( -E_{1,n}/kT \right)/
\sum_{n^{\prime }} \exp\left( -E_{1,n^{\prime }}/kT \right)$ and
$P_{0,m}^{eq} =  \exp\left( -E_{0,m}/kT \right)/
\sum_{m^{\prime}}\exp\left( -E_{0,m^{\prime }}/kT \right)$. Here $\tau $
is the relaxation time which is assumed to be same with
and without a hole in the trap.
The transition rate $W_{\alpha,nm}^{1\rightarrow 0}$ is the rate of hole
tunneling from the state $\left|1,n\right\rangle $ of the trap to
the $\alpha =L, R$ lead
leaving the trap in the state $\left|0,m\right\rangle $. Similarly,
the rate $ W_{\alpha,mn}^{0\rightarrow 1}$ is the rate of hole
tunneling from
the lead $\alpha$ to the state $\left|1,n\right\rangle$ of the
trap, while originally the trap was in the state $\left|0,m\right\rangle$.
These rates can be found from Fermi's golden rule
\begin{equation}
W_{\alpha,nm}^{1\rightarrow 0} =
\Gamma_0 f_{\alpha}\left(E_{1,n}-E_{0,m}\right)
\left|\left\langle 0,m \right|a_{i_0}\left|1,n\right\rangle
\right|^2,
\label{rate1}
\end{equation}
\begin{equation}
W_{\alpha ,mn}^{0\rightarrow 1} = \Gamma_0
\left[ 1 -f_{\alpha } \left( E_{1,n}-E_{0,m}\right) \right]
\left| \left\langle 0,m \right| a_{i_0} \left| 1,n \right\rangle
\right|^2,
\label{rate2}
\end{equation}
where $\Gamma_0=2\pi t_0\rho/\hbar$ and $\rho$ is the density of
states in the leads, which is assumed to be the same in ``L'' and
``R'' leads, and $f_{\alpha }(\epsilon )$ is the Fermi function of the
lead $\alpha $ with a chemical potential $\mu _{\alpha }$.

\begin{figure}
\begin{center}\includegraphics[width=7.6cm]{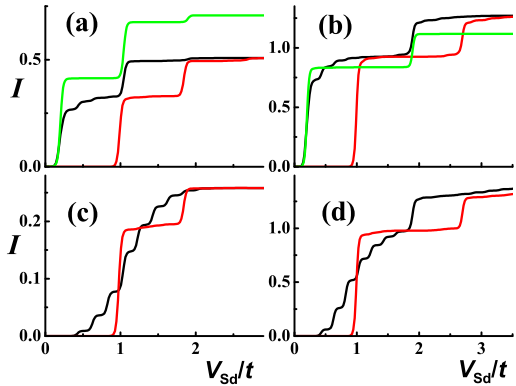}\end{center}
\vspace*{-1cm}
\caption{Current vs the sourse-drain voltage shown for three base pairs
in the trap ($N_t=3$) and $\gamma=0.2$, but for different tunneling
points $i_0$ and  different values of hole-phonon interaction strength:
(a) $i_0=1$: $\lambda =0.5$ and $\epsilon=1.7 t$ (black),
$\epsilon =3.0 t$ (red), $\lambda=0$ and $\epsilon =1.7 t$
(green); (b) $i_0=2$:  $\lambda =0.5$ and $\epsilon =1.7 t$
(black), $\epsilon =3.0 t$ (red), and
$\lambda =0$, $\epsilon =1.7 t$ (green);
(c) $i_0=1$:  $\lambda =1$ and $\epsilon =1.7 t$ (black),
$\epsilon =3.0 t$ (red), (d) $i_0=2$: $\lambda =1$ and
$\epsilon =1.7 t$ (black), $\epsilon =3.0 t$ (red).
}
\label{fig:Fig2}
\end{figure}

For the stationary case the time derivatives of $P_{1,n}$ and
$P_{0,m}$ are zero and Eqs.~(\ref{P1})-(\ref{P0}) becomes the
system of linear equations with condition of normalization
$\sum_{n} P_{1,n}+\sum_{m} P_{0,m}=1$. Then the stationary current
can be calculated as
\begin{equation}
I = \sum_{n,m} \left[  P_{0,m} W_{L, mn}^{0\rightarrow 1} -
P_{1,n} W_{L,nm}^{1\rightarrow 0} \right] .
\label{current}
\end{equation}
We have calculated the current (\ref{current}) numerically as
a function of $V_{sd}$ for different
values of on-site energy, $\epsilon$, which can be changed by the
gate voltage or by doping. By varying $V_{sd}$ we are keeping the
on-site energy $\epsilon $ the same and vary the chemical
potentials of leads as $\mu _L = V_{sd}/2 $ and $\mu _R = -V_{sd}/2$.

To determine the tunneling current we first calculate the
energy spectra of the DNA Hamiltonian [Eq.~(\ref{Htrap})] without
holes and with a single hole in the trap. The $H_{trap}$ then
is just the Hamiltonian of free phonons at each site of the trap.
With a single hole we make the system finite by introducing limitations
on the total number of phonons \cite{marsiglio95} $\sum_i n_{ph,i}\leq10$,
where $n_{ph,i}$ is the number of phonons on site $i$. The
eigenfunctions and eigenvalues of ${\cal H}_{trap}$
can then be found numerically and the corresponding transition rates
Eqs.~(\ref{rate1})-(\ref{rate2}) can be calculated. Finally, we
solve the system of linear eqiations (\ref{P1})-(\ref{P0}) for
a given bias voltage, and substitute this solution into
Eq.~(\ref{current}) to find the stationary tunneling current.

There are five  dimensionless parameters which characterize the I-V structure:
the nonadiabaticity parameter \cite{magna96} $\gamma=\hbar\omega/t$, with
a typical value of $\sim0.01-0.5$ for DNA, the canonical hole-phonon coupling
constant \cite{magna96} $\lambda=\chi^2/(2\hbar\omega t)$ which is
$\sim 0.2-1$ for DNA, dimensionless bias voltage $V_{sd}/t$, on-site energy
$\epsilon/t$, and the ratio of the relaxation time and the tunneling time
$\tau\Gamma_0$.

\begin{figure}
\begin{center}\includegraphics[width=7.6cm]{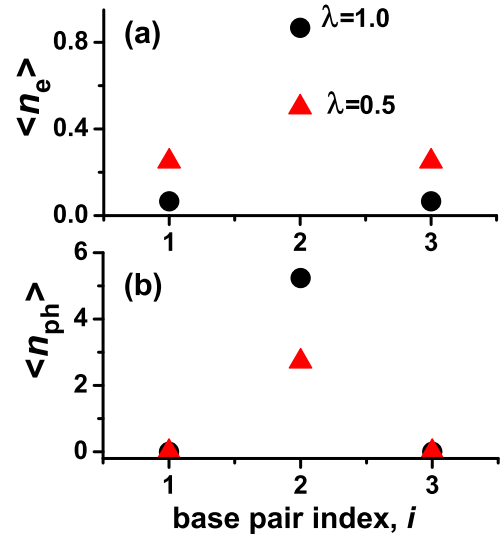}\end{center}
\vspace*{-1cm}
\caption{The average number of holes (a) and
the average number of phonons (b) for a single hole
system in a GGG trap are shown as a function of the base index.
Dots and triangles corresponds to hole-phonon interaction strength
$\lambda =1.0$ and 0.5 eV respectively.
}
\label{fig:Fig3}
\end{figure}

The calculations have been performed for $N_t =2$ and $N_t=3$, i.e.,
for 2 and 3 base pairs in the trap. The example of such a system could be
the guanine hole traps: GG and GGG spots surrounded by adenines. In all the
calculations we kept the ratio of relaxation and tunneling time equal to 1
($\tau \Gamma_0=1$), i.e., the hole-phonon system in the trap is not in the
equilibrium. Different values of $\tau\Gamma_0$, ranging from $\tau\Gamma_0\ll 1$
(equlibrium case) to $\tau\Gamma_0\gg 1$ (unequlibrium case) do
not modify qualitatively the behavior of the I-V curve. To detect the
phonon steps in the I-V curve the temperature should be less then the
phonon frequency and so we keep the temperature equal to $0.01t$.

In Fig.~1 our results are shown for two base
pairs (sites) in the trap. The tunneling occurs through one
of the site, $i_0 = 1$. For uncoupled hole-phonon system
the I-V dependence has two steps corresponding to two single
hole energy levels. The distance between the steps is
$\delta V_{sd} = 2t$. For a small hole-phonon coupling
constant $\lambda =0.5$ [Figs.~1 (a,c)] the additional structures
of width $\delta V_{sd}\simeq\hbar\omega$ due to
the phonon sidebands appear only at the first step and the second
step can still be clearly distinguished. At the same time for a large
gate voltage (large on-site energy $\epsilon $), the phonon
steps are suppressed and the I-V structure becomes similar to that of
a zero-coupling strength, which is shown in Figs.~1 (a,c) by
red lines. For a strong hole-phonon interaction ($\lambda =
1$) the phonon steps suppress the steps due to inter-site
hopping within the trap [Figs.~1 (b,d)]. This suppression becomes
stonger for a larger non-parabolicity $\gamma$, which is illustrated
in Figs.~1 (b,d) by a black line for $\gamma=0.1$ and $\gamma=0.2$.
With increasing gate voltage the phonon steps disappear and the I-V
curve shows a clear two-step structure. The origin of such a suppression
of phonon steps is the following. If the relaxation time is much
smaller than the tunneling time then before tunneling the
hole-phonon system will be in equilibrium. At low temperatures this
means that the system will be at the ground state, i.e., without
any phonons if there are no holes in the trap and in the polaronic
state when there is one hole. Then tunneling from the ``L''
lead will be tunneling to the state without any phonons. Therefore
this process will probe only a single hole state, i.e.
two step structure due to hole hopping. Tunneling to the
``R'' lead is from the ground state of the coupled
hole-phonon system and so this tunneling will produce the phonon
steps in the I-V dependence. Then if the gate voltage or the on-site
energy is increased the tunneling to the ``R'' lead will become
less sensitive to the internal structure of the hole-phonon state
and hence the phonon steps would be suppressed.
From Fig.~1 we conclude that for typical parameters of the DNA structure
the hopping integral between the sites within the DNA traps and phonon
frequency which determine the energetics of hole-phonon
trap system, can be found from dependence of the tunneling current
on $V_{sd}$. From a small gate voltage the phonon
frequency can be found from the I-V curve, while for a larger
gate voltage the hopping integral can be obtained.

The I-V curve should show even richer structure for larger number of
sites in the trap. In Fig.~2 the current as a function of bias voltage
is shown for $N_t=3$ sites. In this case tunneling is possible
through the sites $i_0=1$ and $i_0=2$. For uncoupled
hole-phonon system the I-V curve shows three steps for $i_0=1$
(green line in Fig.~2a), and two steps for $i_0 = 2$ (green line in
Fig.~2b). This means that for $i_0=2$ only two states have non-zero
amplitude at $i=2$ and contribute to the tunneling current. The finite
hole-phonon coupling results in two effects: the phonon steps in
the I-V dependence similar to two-site trap (Fig.~1), and the
polaronic effect which redistributes the hole density along
the trap and increase or decrease the tunneling current.

For small hole-phonon coupling ($\lambda =0.5$) the phonon steps are seen
only at the first hopping step [Fig.~2a,b (black lines)]. The separation
between the steps is the phonon frequency. Similar to Fig.~1, an increase
of the gate voltage (on-site energy) suppresses the phonon steps and the
I-V curve becomes similar in structure to the uncoupled case
[Fig.~2a,b (red lines)]. For a larger hole-phonon coupling
($\lambda=1$) the steps due to hole hopping almost completely disappear
for $i_0 = 1$ [Fig.~2c, black line], but some structure is still
visible for $i_0=2$ [Fig.~2d, black line]. For a larger gate voltage the
steps due to the hopping overcome the phonon effects.

The polaronic effects due to hole-phonon coupling can be clearly
seen in Fig.~2. With incresing hole-phonon interaction the hole
states become more localized at the center of the trap (see Fig.~3)
which results in an increase of the current for tunneling through
the central site of the trap $i_0=2$ (see Fig.~2b,d), and decrease of
the current through $i_0=1$ (see Fig.~2a,c). In addition to changes
to the tunneling current the polaronic effect also modifies the
structure of the I-V curve. This can be seen in Fig.~2a, where with
increasing hole-phonon interaction the third step due to the hole
hopping disappears.

The work of T.C. has been supported by the Canada 
Research Chair Program and the Canadian Foundation for Innovation 
Grant.

\end{document}